\documentclass[iop]{emulateapj}

\newcommand{\src}{\mbox{SN\,2011dh}}

\def\nrao{1}
\def\cfa{2}
\def\bonn{3}
\def\uva{6}
\def\su{7}
\def\york{4}
\def\hart{5}

\begin{document}

\title{EVLA Observations of the Radio Evolution of \src}

\author{M.~I. Krauss\altaffilmark{\nrao}, A.~M. Soderberg\altaffilmark{\cfa}, L. Chomiuk\altaffilmark{\cfa,\nrao}, B.~A. Zauderer\altaffilmark{\cfa}, A.~Brunthaler\altaffilmark{\bonn,\nrao}, M.~F. Bietenholz\altaffilmark{\york,\hart}, R.~A. Chevalier\altaffilmark{\uva}, C. Fransson\altaffilmark{\su}, M. Rupen\altaffilmark{\nrao}}

\altaffiltext{\nrao}{National Radio Astronomy Observatory, Socorro, NM 87801, USA}
\altaffiltext{\cfa}{Harvard-Smithsonian Center for Astrophysics, 60 Garden St., Cambridge, MA 02138, USA}
\altaffiltext{\bonn}{Max-Planck-Institute f\"ur extraterrestrische Physik, Giessenbachstra{\ss}e,  85748 Garching, Germany}
\altaffiltext{\york}{Dept. of Physics and Astronomy, York University, Toronto, M3J 1P3, Ontario, Canada}
\altaffiltext{\hart}{Hartebeesthoek Radio Observatory, P.O. Box 443, Krugersdorp, 1740, South Africa}
\altaffiltext{\uva}{University of Virginia, Astronomy Department, Charlottesville, VA 22904, USA}
\altaffiltext{\su}{Department of Astronomy, The Oskar Klein Centre, Stockholm University, 106 91 Stockholm, Sweden}


\begin{abstract}

We report on Expanded Very Large Array (EVLA) observations of the Type~IIb supernova~2011dh, performed over the first 100~days of its evolution and spanning 1--40~GHz in frequency.  The radio emission is well-described by the self-similar propagation of a spherical shockwave, generated as the supernova ejecta interact with the local circumstellar environment.  Modeling this emission with a standard synchrotron self-absorption (SSA) model gives an average expansion velocity of $v \approx 0.1c$, supporting the classification of the progenitor as a compact star ($R_{\star} \approx 10^{11}\rm~cm$).  We find that the circumstellar density is consistent with a $\rho \propto r^{-2}$ profile.
We determine that the progenitor shed mass at a constant rate of $\approx 3 \times 10^{-5}~M_{\sun}\rm~yr^{-1}$, assuming a wind velocity of $1000\rm~km~s^{-1}$ (values appropriate for a Wolf-Rayet star), or $\approx 7 \times 10^{-7}~M_{\sun}\rm~yr^{-1}$ assuming $20\rm~km~s^{-1}$ (appropriate for a yellow supergiant [YSG] star).  Both values of the mass-loss rate assume a converted fraction of kinetic to magnetic energy density of $\epsilon_B = 0.1$.  
Although optical imaging shows the presence of a YSG, the rapid optical evolution and fast expansion argue that the progenitor is a more compact star---perhaps a companion to the YSG.
Furthermore, the excellent agreement of the radio properties of \src\ with the SSA model implies that any YSG companion is likely in a wide, non-interacting orbit.
\end{abstract}

\section{Introduction}

Type IIb supernovae (SNe~IIb) were first identified as a distinct class of core-collapse events after detailed observations of the ``canonical'' Type IIb SN\,1993J revealed broad hydrogen {\it and} helium features \citep{filippenko97}. Recent studies have shown that this spectroscopic class shows a broad diversity in properties including H$\alpha$ strength, profile, and evolution (e.g., compare with SN\,2003bg; \citealt{matheson,hamuy,mazzali}).  In several cases, YSG stars with extended radii, $R_*\sim 100~R_{\odot}$, have been identified at the explosion sites of SNe~II (1993J, 2008cn, 2009kr, 2011dh; \citealt{93J,maund11,vandyk11,eliasrosa09,eliasrosa10}).  Yet progenitor diagnostics from multi-wavelength studies indicate that some SNe IIb bear stronger similarity to hydrogen-poor Type Ibc supernovae commonly associated with compact progenitors, $R_*\approx R_{\odot}$ (hereafter SNe~cIIb; \citealt{chevalier10}).  In particular, radio-derived estimates for the shockwave velocities are typically high, $v\gtrsim 0.1c$, and difficult to explain in the context of shock breakout from an extended object \citep{nakar10}.  Furthermore, stellar evolution tracks place YSGs outside of the SN explosion phase space of the HR diagram (\citealt{meynet05}, but see \citealt{georgy11}).  In addition, \citet{chevalier10} find that all proposed SNe~cIIb for which there is sufficient radio data show light curve variations indicative of density modulations in the explosion environment, consistent with wind variability from a compact progenitor.  Finally, binary companions have been reported for two SNe~IIb to date (1993J and 2001ig; \citealt{woosley94,ryder06,maund07}).  Similarly, the YSG may be a binary companion rather than the progenitor star.

\begin{deluxetable*}{rlrccccccc}
\tablewidth{0pt}
\tabletypesize{\scriptsize}
\tablecaption{1--8~GHz Radio Flux Densities (mJy)\tablenotemark{a}}
\tablehead{
\colhead{} & \colhead{} & \colhead{} & \multicolumn{6}{c}{Frequency (GHz)} \\
\cline{4-9}
\colhead{Date} & \colhead{MJD} & \colhead{Day\tablenotemark{b}} & \colhead{1.4} & \colhead{1.8} & \colhead{2.5} & \colhead{3.5} & \colhead{4.9} & \colhead{6.7}}
\startdata
Image rms\tablenotemark{c} & \nodata & \nodata & 0.045 & 0.037 & 0.033 & 0.026 & 0.026 & 0.019 \\
June 17 & 55729.2 &  16.4 & \nodata & \nodata & \nodata & \nodata & 2.430$\pm$0.044 & 4.090$\pm$0.063 \\
June 21 & 55733.2 &  20.4 & $<0.13$ & $<0.12$ & 0.540$\pm$0.079 & 1.400$\pm$0.055 & 3.150$\pm$0.043 & 4.800$\pm$0.055 \\
June 26 & 55738.2 &  25.4 & 0.243$\pm$0.079 & 0.800$\pm$0.066 & 1.626$\pm$0.070 & 2.920$\pm$0.061 & 4.920$\pm$0.063 & 5.980$\pm$0.070 \\
July 6 & 55748.1 &  35.3 & 0.331$\pm$0.072 & 1.236$\pm$0.067 & 2.982$\pm$0.073 & 4.908$\pm$0.072 & 6.871$\pm$0.091 & 7.222$\pm$0.078 \\
July 16 & 55758.1 &  45.3 & 0.719$\pm$0.074 & 1.858$\pm$0.069 & 4.092$\pm$0.083 & 6.188$\pm$0.080 & 7.836$\pm$0.086 & 6.987$\pm$0.077 \\
July 29 & 55771.0 &  58.2 & 2.47$\pm$0.12 & 3.31$\pm$0.11 & 5.84$\pm$0.15 & 7.33$\pm$0.14 & 7.47$\pm$0.12 & 6.11$\pm$0.11 \\
Sept 1 & 55805.7 &  92.9 & 3.45$\pm$0.11 & 5.00$\pm$0.12 & 7.02$\pm$0.17 & 6.98$\pm$0.13 & 4.884$\pm$0.073 & 3.941$\pm$0.061 \\
\enddata
\label{srcTab1}
\tablenotetext{a}{Quoted upper limits are $3\sigma$.}
\tablenotetext{b}{Days from 1 June 2011 (MJD 55712.8).}
\tablenotetext{c}{Mean image rms for the first five epochs.  The final two epochs have rms values $\approx  \sqrt{8}$ times higher, since the available bandwidth was 1/8 that of previous observations.}
\end{deluxetable*}

\begin{deluxetable*}{rlrccccccc}
\tablewidth{0pt}
\tabletypesize{\scriptsize}
\tablecaption{8--37~GHz Radio Flux Densities (mJy)\tablenotemark{a}}
\tablehead{
\colhead{} & \colhead{} & \colhead{} & \multicolumn{7}{c}{Frequency (GHz)} \\
\cline{4-10}
\colhead{Date} & \colhead{MJD} & \colhead{Day\tablenotemark{b}} & \colhead{8.4} & \colhead{13.5} & \colhead{16.0} & \colhead{20.5} & \colhead{25.0} & \colhead{29.0} & \colhead{36.0}}
\startdata
Image rms\tablenotemark{c} & \nodata & \nodata & 0.015 & 0.026 & 0.029 & 0.034 & 0.036 & 0.028 & 0.037 \\
June 17 & 55729.2 &  16.4 &  5.535$\pm$0.057 &  6.970$\pm$0.074 &  6.790$\pm$0.073 &  6.50$\pm$0.20 &  5.13$\pm$0.16 &  4.60$\pm$0.14 &  3.47$\pm$0.11 \\
June 21 & 55733.2 &  20.4 &  5.870$\pm$0.060 &  5.940$\pm$0.064 &  5.313$\pm$0.057 &  4.56$\pm$0.14 &  3.61$\pm$0.12 &  3.117$\pm$0.097 &  2.190$\pm$0.074 \\
June 26 & 55738.2 &  25.4 &  6.935$\pm$0.071 &  5.574$\pm$0.080 &  4.744$\pm$0.096 &  3.70$\pm$0.13 &  2.88$\pm$0.11 &  2.349$\pm$0.077 &  1.644$\pm$0.064 \\
July 6 & 55748.1 &  35.3 &  6.820$\pm$0.071 &  4.334$\pm$0.068 &  3.917$\pm$0.073 &  2.92$\pm$0.11 &  2.53$\pm$0.10 &  2.063$\pm$0.073 &  1.772$\pm$0.074 \\
July 16 & 55758.1 &  45.3 &  6.082$\pm$0.064 &  3.790$\pm$0.063 &  2.960$\pm$0.067 &  2.493$\pm$0.097 &  1.819$\pm$0.080 &  1.549$\pm$0.053 &  1.159$\pm$0.051 \\
July 29 & 55771.0 &  58.2 &  5.097$\pm$0.057 &  2.83$\pm$0.14 &  2.32$\pm$0.14 &  2.35$\pm$0.15 &  1.53$\pm$0.16 &  1.32$\pm$0.15 & $<  0.69$ \\
Sept 1 & 55805.7 &  92.9 &  2.891$\pm$0.043 &  1.627$\pm$0.064 &  1.321$\pm$0.072 &  1.28$\pm$0.15 &  0.71$\pm$0.10 &  0.60$\pm$0.11 & $<  0.42$ \\
\enddata
\label{srcTab2}
\tablenotetext{a}{Quoted upper limits are $3\sigma$.}
\tablenotetext{b}{Days from 1 June 2011 (MJD 55712.8).}
\tablenotetext{c}{Mean image rms for the first five epochs.  The final two epochs have rms values $\approx  \sqrt{8}$ times higher, since the available bandwidth was 1/8 that of previous observations.}
\end{deluxetable*}

Probing the distinguishing differences between SNe~cIIb and extended Type~IIb supernovae requires early discovery, since rapid follow-up observations are crucial for identifying unique characteristics.
Since the advent of dedicated transient surveys and improvements in amateur astronomical equipment, such 
early discoveries are becoming more common.  In June 2011, an optical transient was found in M51 by amateur astronomer Am\'ed\'ee Riou \citep{iaucriou}.  Prompt spectroscopic
follow-up indicated that it was a Type II supernova, and further spectroscopy revealed that the object most closely
resembled a Type IIb \citep{arcavi11,marion11}.  A YSG was identified
in pre-explosion HST imaging at the position of the SN
\citep{maund11,vandyk11}, similar to the case of SN\,1993J.  However, 
rapid optical follow-up observations from the Palomar Transient
Factory \citep{law} pointed to a compact progenitor star, as evidenced by
its short-lived ($\Delta t\approx 1$~day) cooling-envelope emission \citep{arcavi11}.
In our recent paper (\citealt{soderberg11}; hereafter Paper~I) we
reported on early radio, mm-band and X-ray emission. 
Based on modeling of the non-thermal emission, we found that the
shockwave velocity was $v\sim 0.1c$, more typical of a SN~Ibc (assumed to have a compact Wolf-Rayet progenitor) than the explosion of an extended supergiant.  If \src\ did, in fact, have a YSG progenitor, it would be necessary to explain the high shockwave velocity and rapid optical evolution in the context of an extended star.

In this paper, we present our detailed Expanded Very Large Array (EVLA; \citealt{perley11})
observations of \src\ spanning $\Delta t\approx 100$ days after
explosion.  This project capitalizes on the nearly continuous coverage now available from 1--40~GHz with the EVLA.  We model the synchrotron emission over 7 epochs to derive the evolution of the
shockwave radius and magnetic field as a function of time. We confirm
the initial results of Paper I and find that the radio emission over the course of these observations
evolves smoothly, with no evidence as yet for the circumstellar density variations seen in other compact SNe~IIb.

\section{Observations and Data Reduction}\label{observations}

We obtained multi-frequency monitoring observations with the EVLA
beginning 17 days after explosion (taking $t_0 =$~2011 May 31.8 UTC) and continuing through day 92 (Program 11A-277: PI Soderberg). These observations comprise seven epochs with roughly logarithmic spacing, matching the expected evolution of the supernova light
curves.  The second through final epochs covered 1.0--36.5~GHz,
utilizing the L (1--2~GHz), S (2--4~GHz), C (4--8~GHz), X (8--8.8~GHz), Ku (12--18~GHz), K (18--26.5~GHz), and Ka-band (26.5--40~GHz) receivers (see \citealt{perley11} for a description of new observing capabilities with the EVLA).  Within each observing band (except X-band),\footnote{Very few wideband X-band receivers were available at the time of our observations.  The total bandwidth at X-band was 0.8~GHz and 256~MHz.} each of two basebands was tuned to a different frequency in order to maximize
spectral coverage.  We took data with the maximum available 1~GHz of bandwidth
per baseband at all bands during the first five epochs, and 128~MHz
for the final two. Each epoch was three hours in duration.  All observations were performed while the EVLA was in its most
extended A-configuration, giving the highest available spatial
resolution.  EVLA observations were not continued in the
subsequent (most compact) D-configuration, due to concerns about
confusion with other sources of radio emission in M51 at lower frequencies, to which the spectral peak of \src\ had shifted by this time.

Phase reference calibration was carried out using J1335+4542
(1.7\degr~away; 8--40~GHz) or J1327+4326 (3.8\degr~away; 1--8~GHz).  At the highest frequencies ($>20$~GHz), rapid switching (2~minute cadence) between \src~and J1335+4542 was done  to enable
optimal correction for atmospheric phase variations, and referenced pointing was used to correct for antenna pointing offsets.  Each observation included data at all bands for the standard flux density calibration source
3C286.  Processing was performed with NRAO's Common Astronomy Software
Applications (CASA; \citealt{mcmullin07}) or Astronomical Image Processing System (AIPS; \citealt{greisen03}), using the same
procedure in each.

Bad data identified by the EVLA online system were deleted, as were pure zeros (sometimes generated by the correlator as a result of failure); further editing out of radio-frequency interference and poorly performing antennas was done by hand.  
Frequency-dependent atmospheric opacity was accounted for using the average of a seasonal model and observation-specific information from the weather station \citep{marvil10}.  At frequencies higher than 5~GHz, where elevation-dependent antenna gain effects become important ($\gtrsim 1\%$~variance from zenith values), gain curve information was applied at relevant points in the calibration process.

\begin{figure*}
\resizebox{\textwidth}{!}{\includegraphics{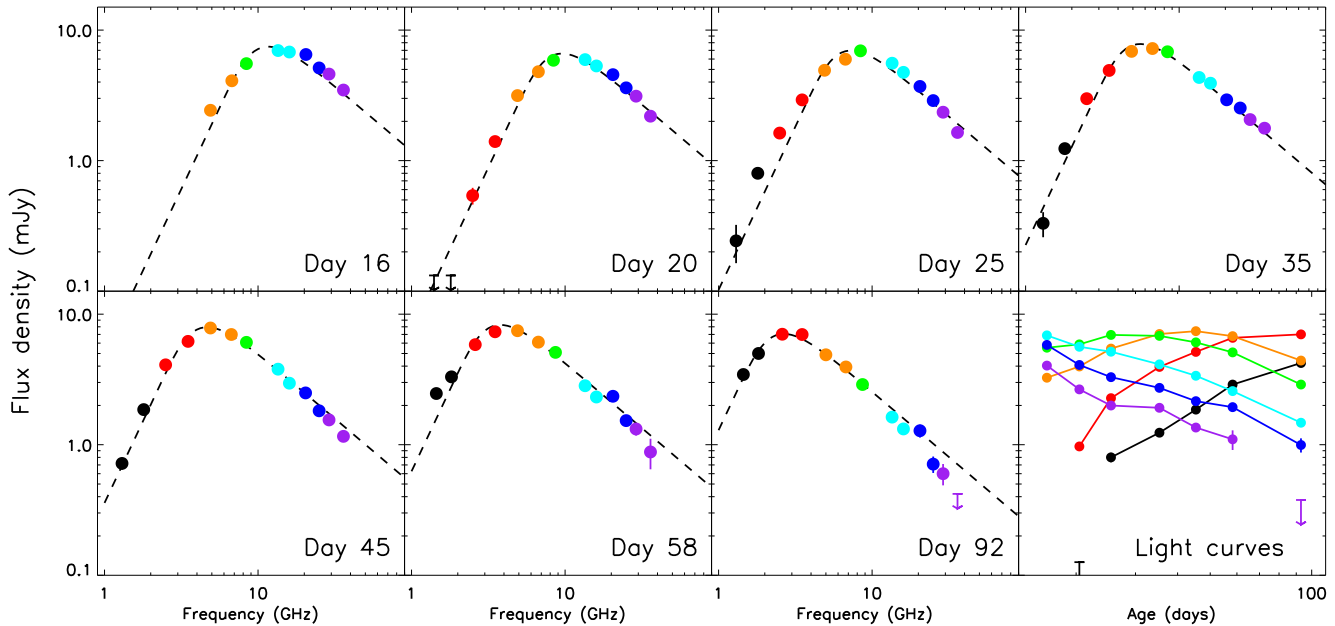}}
	\caption{{\bf Panels 1$-$7:} Spectra of \src.  Dashed curves show corresponding SSA model fits.  In addition to errors derived from Gaussian source fitting and image rms, we include 1\% and 3\% systematic errors at low ($< 20$~GHz) and high ($> 20$~GHz) frequencies; these are generally smaller than the plotted symbols. {\bf Panel 8:} Light curves of \src.  Data are averaged for each receiver band and color-coded as in the preceding spectral plots.}
	\label{specPlot}
\end{figure*}

To calibrate the data, a bandpass solution was derived using 3C286; applying this solution, we solved for the complex gains of the calibration sources.  We scaled the amplitude gains of the phase calibrator according to the values derived for 3C286 using the ``Perley-Butler 2010'' flux density standard, and applied these solutions to \src.\footnote{Due to a problem with data acquisition, the flux density calibrator was not present at Ku-band for epoch~3.  We characterized the variability of J1335+4542 over the other epochs and found it to be small (rms of approximately 1\% at both 13.5~and 16.0~GHz).  We fit its average spectrum and used this model to flux-calibrate the data for \src, adding the additional source of error from J1335+4542's variability in quadrature with the image noise and fit errors to determine the total uncertainty on flux density.}  When there was sufficient signal, we performed phase-only self-calibration, and natural weighting was used during the imaging process.  We fit an elliptical Gaussian model at each frequency to derive integrated flux density values.  To estimate the uncertainties on our measured flux densities, we added (in quadrature) uncertainties from the Gaussian fitting with the rms noise of the images, as well as 1\% systematic errors at low frequencies ($<20$~GHz) and 3\% at high frequencies ($>20$~GHz).  The integrated flux density values and associated 1$\sigma$ uncertainties, as well as mean rms image noises for each frequency, are reported in Tables~\ref{srcTab1} and~\ref{srcTab2}.

\section{Modeling and Results}

Radio emission from supernovae arises when the expanding ejecta interact with pre-existing circumstellar material, which for SN~IIb is provided by the progenitor's stellar wind.  
The interaction of the supernova blast wave with the circumstellar environment probes recent stellar mass-loss \citep{chevalier06}.  Here, we model the radio emission from \src\ using the standard circumstellar interaction model (``model 1'' in \citealt{chevalier96}): as the expanding shock moves into the circumstellar medium, the magnetic field strength in the interaction zone increases via turbulence generated in the shocked region.
Electrons that have been accelerated to relativistic energies interact with this enhanced magnetic field, producing synchrotron emission.  This emission is subject to self-absorption, as further explored by \citet{chevalier98}; we employ this form of the synchrotron self-absorption (SSA) model to provide an analytic description of the observed radio spectra.  

\subsection{SSA Model Fits and Derived Parameters}\label{analysis}

For each given epoch, we fit the radio spectrum using the parameterization

\begin{equation}
S(\nu) = 1.582 \, S_{\nu_{\tau}} \left(\frac{\nu}{\nu_{\tau}}\right)^{5/2} \left\{1-\exp\left[-\left(\frac{\nu}{\nu_{\tau}}\right)^{-(p+4)/2}\right]\right\},
\end{equation}

\noindent where $S_{\nu_{\tau}}$ is the flux density at $\nu_{\tau}$, the frequency at which the optical depth is unity, and
$p$ the electron power-law index \citep{chevalier98}.  We do not see any evidence of external free-free absorption, as was found for SN\,1993J \citep{weiler07}.  Including free-free absorption in our fits did not reduce the resulting $\chi^2$ values; the fitting minimized its effect to an insignificant contribution. 
This is not surprising, since the spectra often show excess emission relative to the SSA model at low frequencies---the opposite of what is expected if free-free absorption were significant. 

Allowing $p$ to vary did not significantly improve the resulting reduced $\chi^2$ values, so we froze this parameter to its average fitted value of $p = 2.8$ to limit the number of free parameters.  For $p = 2.8$, the observed peak radio flux $S_{\nu_{\rm op}}$ occurs at $\nu_{\rm op} = 1.17\,\nu_{\tau}$.

Although the SSA model likely represents an over-simplification of the actual source geometry and physics, it provides reasonable fits to the observations (see Figure~\ref{specPlot}).  Despite some systematic deviations, the model matches the peaks in the spectra well; large variations of the true average radii from the ones implied by the models therefore seem unlikely.  

However, the formal errors described in \S\ref{observations} are small relative to deviations from the fits, resulting in large values of $\chi^2$ and unacceptably small uncertainties on the fitted values of  $S_{\nu_{\tau}}$ and $\nu_{\tau}$.  This is probably due to a combination of underestimated formal errors (the EVLA was still in its commissioning phase at the time of the observations), as well as unaccounted for physics that is beyond the scope of this paper.  To better estimate the errors on the fitted parameters, we scaled the fitted uncertainties for each epoch so that the reduced $\chi^2$ values were unity, equivalent to increasing the errors by a factor of 3--7.  These uncertainties were propagated throughout subsequent calculations.  Values for $S_{\nu_{\tau}}$ and $\nu_{\tau}$, as well as the parameters derived below, are presented in Table~\ref{fitResults}.

We find that the SSA-derived peak frequencies are systematically $\approx 10\%$ lower than the apparent peaks, but since this trend is consistent over the course of the observations, it will not affect the time-dependence of the derived parameters.  We also note that for the final three epochs (days 45, 58, and 92), the high-frequency data trend below the model.  This is likely due to the fact that for these epochs, there was insufficient flux to perform self-calibration above 20~GHz (K or Ka-bands), resulting in possible underestimation of the flux densities due to phase decoherence.

The fitted spectra can be used as observational tracers of the outer shock radius ($R_s$), strength of the magnetic field ($B_s$), and density of the progenitor's wind ($\rho_{\rm wind}$), given a minimal set of assumptions \citep{chevalier98,chevalier06}:  

\begin{eqnarray}
R_s &=& 3.9 \times 10^{14} \, \alpha^{-1/19} \left( \frac{f}{0.5} \right)^{-1/19}  \left( \frac{D}{\rm Mpc} \right)^{18/19} \nonumber \\ 
 & & \left( \frac{S_{\nu_{\rm op}}}{\rm mJy} \right)^{9/19} \times \left( \frac{\nu_{\rm op}}{\rm 5~GHz} \right)^{-1} {\rm cm}, \\
B_s &=& 1.0 \, \alpha^{-4/19} \left( \frac{f}{0.5} \right)^{-4/19} \left( \frac{D}{\rm Mpc} \right)^{-4/19} \nonumber \\
 & & \times \left( \frac{S_{\nu_{\rm op}}}{\rm mJy} \right)^{-2/19} \left( \frac{\nu_{\rm op}}{\rm 5~GHz} \right) {\rm G},
\end{eqnarray}

\noindent and

\begin{eqnarray}
\rho_{\rm wind} &=& A r^{-2} {\rm~g~cm^{-3}},
\end{eqnarray}

\noindent where the circumstellar density is parametrized as
$A_{*} = A/(5 \times 10^{11}\rm~g~cm^{-1})$, and 
\vspace{-0.1in}

\begin{eqnarray}
A_{*} &=& 0.82 \, \alpha^{-8/19} \left( \frac{\epsilon_B}{0.1} \right)^{-1} \left( \frac{f}{0.5} \right)^{-8/19} \left( \frac{D}{\rm Mpc} \right)^{-8/19} \nonumber \\
 & & \times \left( \frac{S_{\nu_{\rm op}}}{\rm mJy} \right)^{-4/19} \left( \frac{\nu_{\rm op}}{\rm 5~GHz} \right)^2 \left( \frac{t}{10\rm~d} \right)^2.
\end{eqnarray}

\begin{figure}
\resizebox{\columnwidth}{!}{\includegraphics{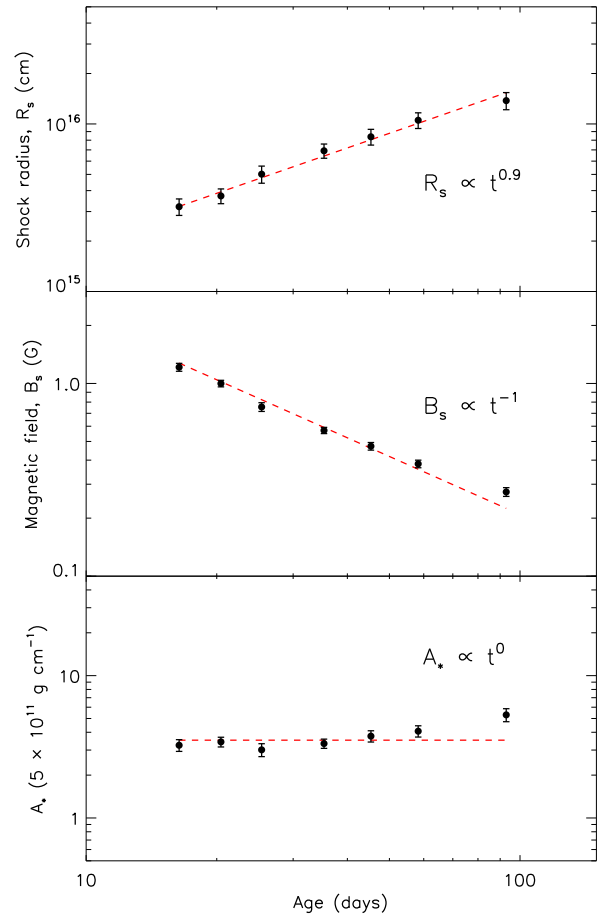}}
	\caption{{\bf Upper panel:} shock radius, {\bf center panel:} magnetic field strength, and {\bf lower panel:} circumstellar density parameter derived from the per-epoch SSA fits.  The expected time behavior from the standard model of \citet{chevalier96} is shown with dashed lines.}
	\label{RpBpPlot}
\end{figure}

\noindent Here, $\alpha$ is the ratio of electron to magnetic energy densites ($u_e / u_B$), $f$ the filling fraction of emitting material, $D$ the distance, $t$ the age, and $\epsilon_B$ the converted fraction of kinetic to magnetic energy density ($\epsilon_B = u_B / \rho_{\rm wind}\,v_s$).  We assume equipartition ($\alpha = 1$), and take $p=2.8$ (as fitted).  In addition, we assume a filling factor $f = 0.5$ (approximately as was found for SN\,1993J; \citealt{bartel02}), and a distance $D = 8.4\pm0.6$~Mpc \citep{feldmeier97,vinko11}.  The time evolution of the shock radius is consistent with $R_s \propto t^{0.9}$ (see also \citealt{bietenholz11}), and the magnetic field strength with $B_s \propto t^{-1}$ (top and center panels of Figure~\ref{RpBpPlot}).  

\begin{deluxetable*}{lccccc}
\tablewidth{0pt}
\tabletypesize{\small}
\tablecaption{SSA model fits}
\tablehead{
\colhead{Day\tablenotemark{a}} & \colhead{$S_{\nu_{\tau}}$ (mJy)} & \colhead{$\nu_{\tau}$ (GHz)} & \colhead{$R_s$ ($10^{15}$ cm)} & \colhead{$B_s$ (G)} & \colhead{$A_{*}$\tablenotemark{b}}}
\startdata
16.4 &  7.03$\pm$0.25 & 10.01$\pm$0.29 &  3.2$\pm$0.4 & 1.21$\pm$0.06 &  3.2$\pm$0.3 \\
20.4 &  6.19$\pm$0.15 &  8.13$\pm$0.18 &  3.7$\pm$0.4 & 1.00$\pm$0.04 &  3.4$\pm$0.3 \\
25.4 &  6.52$\pm$0.22 &  6.18$\pm$0.21 &  5.0$\pm$0.6 & 0.76$\pm$0.04 &  3.0$\pm$0.3 \\
35.3 &  7.28$\pm$0.15 &  4.72$\pm$0.10 &  6.9$\pm$0.7 & 0.57$\pm$0.02 &  3.3$\pm$0.2 \\
45.3 &  7.36$\pm$0.19 &  3.91$\pm$0.11 &  8.4$\pm$0.9 & 0.47$\pm$0.02 &  3.8$\pm$0.3 \\
58.2 &  7.69$\pm$0.20 &  3.184$\pm$0.088 & 11$\pm$1 & 0.38$\pm$0.02 &  4.1$\pm$0.4 \\
92.9 &  6.44$\pm$0.21 &  2.235$\pm$0.076 & 14$\pm$2 & 0.27$\pm$0.01 &  5.3$\pm$0.6 \\
\enddata
\label{fitResults}
\tablenotetext{a}{Days from 1 June 2011 (MJD 55712.8).}
\tablenotetext{b}{$5 \times 10^{-11}\rm~g~cm^{-2}$}
\end{deluxetable*}

The expectation of $\rho_{\rm wind} \propto r^{-2}$ may be questionable at large radii, but it is reasonable to approximate the immediate circumstellar environment by assuming a constant progenitor wind 
\citep{dwarkadas11}.  These observations probe a region extending to $\sim 1000$~AU, corresponding to $\sim 5$~yr for a $1000~\rm km~s^{-1}$ wind --- much shorter than necessary for substantial wind variability.  Given these constraints, and taking $\epsilon_B = 0.1$, we find $A_{*} \approx 3.5$.  There is no strong evidence for time variability, suggesting that our assumption of a constant progenitor wind was reasonable (bottom panel of Figure~\ref{RpBpPlot}).

In Paper~I, joint radio and X-ray model fits pointed to deviations from equipartition, $\alpha \approx 30$ and $\epsilon_B \approx 0.01$, under the assumption that inverse Compton emission dominates the X-ray signal.  With these values, our radius estimates are smaller by a factor of 0.8, magnetic field values smaller by a factor of 0.5, and $A_*$ larger by a factor of 2.

\subsection{Physical Interpretation}\label{discussion}

The derived outer shock radii imply an average shock velocity of $dR_s/dt \approx 25,000$~km~s$^{-1}$, or $\approx 0.1c$, in agreement with Paper~I.  As noted there, the shockwave is traveling a factor of $\sim1.5$ faster than material at the optical photosphere ($\approx 17,000$~km~s$^{-1}$ at $\Delta t \approx 3$~days; \citealt{silverman11}). 

Assuming that the supernova ejecta and progenitor wind have power-law density profiles, the time evolution of the shock radius can be expressed as $R_s \propto t^m$.  For the expected circumstellar density $\rho_{\rm wind} \propto r^{-2}$, $m = (n-3)/(n-2)$, where $n$ is the power-law index of the outer supernova density profile.  
We find $m = 0.87\pm0.07$, corresponding to $n = 9.7^{+12}_{-3.7}$, and reasonable for a fast blastwave from a compact progenitor (canonical value of $m = 0.9$; \citealt{chevalier92}).  This value of $m$ is also consistent with joint EVLA-VLBI fits reported in our companion paper \citep{bietenholz11}.

The measured decrease in magnetic field strength agrees with the standard model for the hydrodynamic evolution of a self-similar shock.   The magnetic field generation is thought to arise via turbulence in the shocked region, so it is proportional to the total post-shock energy density ($\propto t^{-2}$; \citealt{chevalier98}); therefore, $B_s \propto t^{-1}$, as observed.

We find that the scaling factor for the circumstellar density is consistent with a constant value, $A_{*} \approx 3.5$, over the course of our observations.   This implies a constant mass-loss rate of $\approx 3 \times 10^{-5}\rm~M_{\sun}~yr^{-1}$, assuming a wind velocity of $1000~\rm km~s^{-1}$.  These values are in the expected range for a Wolf-Rayet progenitor \citep{crowther07}, and agree with our analysis of the early-time radio data (Paper~I).  Since the implied mass-loss rate scales linearly with wind velocity ($\dot{M} \propto A \, v_{\rm wind}$), a wind velocity of $20\rm~km~s^{-1}$ gives a mass-loss rate of $\approx 7 \times 10^{-7}\rm~M_{\sun}~yr^{-1}$, reasonable for a YSG progenitor \citep{georgy11}.

However, the high shock velocity, as well as the rapid cooling observed in early-time optical spectra, suggest a compact progenitor star for \src, and a Type~cIIb classification (\citealt{arcavi11}, Paper~I).  Furthermore, the fitted electron power-law index of $p=2.8$ is close to what is typically found for observations of SNe~Ibc \citep[$p \approx 3$;][]{chevalier06}, which are presumed to have compact progenitors.  

One further observational characteristic of Type~cIIb~SNe is that they often display late-time radio variability, as was seen in SN\,2001ig and SN\,2003bg \citep{ryder04,soderberg06}.  We do not yet see evidence of variability in \src, but note that our observations only cover the first $\sim 100$~days of evolution, around the time that variability was discovered in other SNe~cIIb.  

\section{Conclusions}

The radio spectra of \src\ are well-characterized by an SSA model without any need for additional free-free absorption.  This is in contrast with SN\,1993J, which required both, and also had a substantially higher derived mass-loss rate ($A_* \approx 500$, compared with $\approx 4$ for \src; \citealt{fransson98}).  SN\,1993J was classified as a SN~eIIb because of its extended progenitor \citep{chevalier10,woosley94}: the observed differences in absorption and circumstellar density could be characteristic of these two classes of IIb~SNe.

However, pre-supernova images of M51 show a YSG co-located with the explosion site, which was suggested as a potential progenitor or binary companion \citep{maund11,vandyk11,murphy11}.  As we have mentioned, it seems unlikely that the YSG is in fact the progenitor.  If it were, this would require a process to enable the escape of fast shockwaves from YSGs, either through steep ejecta density profiles or ejecta asymmetries.

If the YSG was a binary companion, then some interaction of the shock and the YSG might be expected.  The observational agreement of the radio measurements with the standard model suggests that this did not occur within $\sim1000$~AU of the explosion site.  In this scenario, then, the orbit must have been quite wide, with an orbital period 6000~yr, so any interaction between the YSG and the supernova progenitor would have been limited. For comparison, the binary progenitor of SN\,1993J likely had an orbital period of $\sim 2000$~d \citep{stancliffe09}, allowing substantial mass transfer to occur and stripping the presupernova star of its H-rich envelope \citep{woosley94}. This would not have been possible for \src. Alternatively, if the blast wave were highly asymmetric, interaction with a more nearby companion could have been minimized.  We consider this unlikely, since the radio data agree quite well with the standard, spherical ejecta model.  Furthermore, the YSG phase is estimated to last only $\sim 3000$~yr \citep{drout09}, making it improbable that the companion would happen to be a YSG at the time of the supernova.

Finally, it may be that the YSG is unrelated to the supernova, and is only coincidentally along the same line of sight. Future observations, including optical imaging of the field after SN 2011dh has faded, will help determine any association with the YSG and the true nature of the progenitor system.

\acknowledgments{The National Radio Astronomy Observatory is a facility of the National Science Foundation operated under cooperative agreement by Associated Universities, Inc.  A. B. was supported by a Marie Curie Outgoing International Fellowship (FP7) of the European Union (project number 275596).  We thank the anonymous referee for helpful comments and suggestions.  We acknowledge with thanks the variable star observations from the AAVSO International database contributed by observers worldwide and used in this research.  This research has made use of NASA's Astrophysics Data System Bibliographic Services.}

\end{document}